\documentclass{amsart}
\usepackage{amsthm}
\usepackage{amsmath}
\usepackage{amsfonts}
\usepackage{amssymb}
\usepackage{mathrsfs}
\usepackage{hyperref}
\usepackage{fancyhdr}
\usepackage{graphicx}
\usepackage{float}                  
\usepackage{rotating}               
\restylefloat{figure}               
\theoremstyle{plain}
\newtheorem{lem}{Lemma}[section]

\newtheorem{thm}[lem]{Theorem}
\newtheorem{cor}[lem]{Corollary}

\theoremstyle{definition}

\newtheorem{rem}[lem]{Remark}
\def\N{{\mathbb N}}
\def\R{{\mathbb R}}
\def\C{{\mathbb C}}

\def\Z{{\mathbb Z}}

\def\T{{\mathbb T}}
\def\PP{\mathcal P}
\def\NN{\mathcal N}
\def\MM{\mathcal M}
\def\P{\mathcal P}
\def\ee{\text{e}}
\def\eps{\varepsilon}
\def\Cci#1{C_c^\infty(#1)}
\def\Ccinf{C_c^{\infty}}
\def\dx{\,\text{d}x}
\def\d{\,\text{\rm d}}
\def\sigmaess{\sigma_{\text{\rm ess}}}
\def\sigmadisc{\sigma_{\text{\rm disc}}}
\def\supp{\text{supp }}
\def\ska#1#2{\left<#1,#2\right>}
\def\norm#1{\left|\!\left|{#1}\right|\!\right|}
\def\infnorm#1{{\left|\!\left|#1\right|\!\right|}_{\infty}}
\def\dist{\text{dist}}
\def\cl#1{\overline{#1}}
\def\phi{\varphi}
\def\theta{\vartheta}
\begin{document}
\title{A variational approach to dislocation problems for periodic Schr\"odinger operators}
\author{Rainer Hempel}
\author{Martin Kohlmann}
\address{Institute for Computational Mathematics, Technische
Universit\"at Braun\-schweig, Pockelsstra{\ss}e 14, 38106
Braunschweig, Germany}
\email{r.hempel@tu-bs.de}
\address{Institute for Applied Mathematics, Leibniz Universit\"at Hannover, Welfengarten 1, 30167 Hannover, Germany}
\email{kohlmann@ifam.uni-hannover.de}
\begin{abstract} As a simple model for lattice defects like grain
boundaries in solid state physics we consider potentials which are
obtained from a periodic potential $V = V(x,y)$ on $\R^2$ with
period lattice $\Z^2$ by setting $W_t(x,y) = V(x+t,y)$ for $x < 0$
and $W_t(x,y) = V(x,y)$ for $x \ge 0$, for $t \in [0,1]$. For
Lipschitz-continuous $V$ it is shown that the Schr\"odinger
operators $H_t = -\Delta + W_t$ have spectrum (surface states) in
the spectral gaps of $H_0$, for suitable $t \in (0,1)$. We also
discuss the density of these surface states as compared to the
density of the bulk. Our approach is variational and it is first
applied to the well-known dislocation problem [K1, K2] on the real
line. We then proceed to the dislocation problem for an infinite
strip and for the plane. In an appendix, we discuss regularity
properties of the eigenvalue branches in the one-dimensional
dislocation problem for suitable classes of potentials.
\end{abstract}
\maketitle
\section{Introduction}
 In solid state physics, one first studies crystallized matter with a
 perfectly regular atomic structure where the atoms are located on a
 periodic lattice. However, most crystals are not perfectly periodic;
 in fact,
 the regular pattern of atoms may be disturbed by various defects which
 fall into two main classes: there are defects which leave the
 lattice unchanged (like impurities or vacancies), and there are
 more serious ``geometric'' defects of the lattice itself, cf.\ [AM],
 which may involve translations and rotation of portions of the
 lattice. Such lattice dislocations occur, in particular, at grain boundaries in alloys.
 These models are deterministic but may be generalized to include randomness.

 Many of the geometric defects mentioned above are accessible to
 mathematical analysis only after some idealization which leads to
 the following type of problem, cf.\ [DS]: there is a periodic potential $V \colon
 \R^d \to \R$ with period lattice $\Z^d$ and a Euclidean transformation
 $T \colon \R^d \to \R^d$ such that the potential coincides with $V$
 in the half-space $\{ x \in \R^d \mid x_1 \ge 0\}$ and with $V \circ T$ in $\{x_1 < 0\}$.
 In the simplest cases $T$ is translation in the direction
 of one of the coordinate axes, with again two main subcases:
 translation orthogonal to the hyperplane $\{ x_1 = 0 \}$ or translations that
 keep the $x_1$-coordinate fixed.
 In the present paper, we discuss the case $d=2$ (where the coordinates are denoted by
 $x$ and $y$) and we will mainly focus on translation in the $x$-direction.
 In a forthcoming companion paper [HK] we will then study
 some aspects of the {\it rotation problem} where we take the given periodic potential
 $V$ in the right half-plane and a rotated version $V \circ M_\theta$
 in the left half-plane with $M_\theta$ denoting rotation by the angle
 $\theta$; some results from the present paper will be essential for [HK].

 The one-dimensional dislocation problem is particularly simple:
 Let $V:\R\to\R$ be a periodic potential with period $1$ and let
 $$
 W_t(x):=\left\{%
 \begin{array}{lll}
  V(x),   && x\geq 0,\\
  V(x+t), && x<0,\\
 \end{array}%
 \right.\eqno{(1.1)}
 $$
 for $t\in[0,1]$. The (self-adjoint) operator  $H_t:=-\frac{\text{d}^2}
 {\text{d}x^2} + W_t$ is called the {\it dislocation operator},
 $t$ the {\it dislocation parameter}. There is quite a number of results
 available on this problem:
 it is well known and easy to see that the essential spectrum of $H_t$
 does not depend on $t$ for $0 \le t \le 1$; also $H_t$
 cannot have any embedded eigenvalues.
 Furthermore, there is no singular continuous spectrum, cf.\ [DS].
 For $0 < t < 1$,  the operators $H_t$ may have bound states (discrete eigenvalues)
 located in the gaps of the essential spectrum. These eigenvalues and
 the corresponding resonances have been studied by Korotyaev
 [K1, K2] in great detail, using powerful results from analytic function
 theory which are specific to the one-dimensional, periodic case.
 While, predictably, our results for the one-dimensional periodic case are weaker
 than Korotyaev's, our method of proof is very elementary and can
 be generalized in several directions; most importantly, we can
 apply our techniques to dislocation problems in dimensions greater than 1.
 In one dimension, we also give a more systematic treatment of
 regularity properties of the eigenvalue ``branches''; in particular,
 it is shown that the eigenvalue branches are Lipschitz-continuous if $V$
 is (locally) of bounded variation.

 The one-dimensional dislocation problem is mainly included to
 introduce and test our variational approach which is inspired by [DH, ADH]:
 we use approximations  by problems on intervals $(-n-t,n)$ with
 periodic boundary conditions where it is easy to control the spectral flow, and let $n$ tend to
 $\infty$. This  idea can be adapted to the study of the translation problem
 for the strip $\Sigma := \R \times (0,1)$ in $\R^2$ with periodic
 boundary conditions in the $y$-variable, say. In $\R^2$, we consider
 dislocation potentials $W_t$ defined by
 $$
 W_t(x,y):=\left\{%
 \begin{array}{lll}
  V(x,y),   && x\geq 0,\\
  V(x+t,y), && x<0,\\
 \end{array}%
 \right.\eqno{(1.2)}
 $$
 for $t \in [0,1]$. On the strip $\Sigma$ we obtain
 existence results for eigenvalues of $S_t := -\Delta + W_t$ in the spectral
 gaps of $S_0$. From that we easily derive that $D_t := -\Delta + W_t$, acting in $L_2(\R^2)$,
 will have {\it surface states} with a non-zero
 density on an appropriate scale, for suitable $t \in (0,1)$.
 To distinguish the bulk from the surface density of states for this problem,
 we consider the operators $-\Delta + W_t$ on squares
  $Q_n = (-n,n)^2$  with Dirichlet boundary conditions, for
 $n$ large, count the  number of eigenvalues inside a compact
 subset of a non-degenerate spectral gap of $D_0$ and scale with $n^{-2}$
 for the bulk and  with $n^{-1}$ for the surface states. Taking the limits $n\to\infty$
 (which exist as explained in [DS, EKSchrS]), we obtain
 the integrated density of states measures
 $\rho_{\text{bulk}}(D_t,I)$ for the bulk and
 $\rho_{\text{surf}}(D_t,J)$ for the surface states of this model;
 here $I \subset \R$ and $J \subset \R\backslash\sigma(D_0)$ are open intervals and
 $\overline J \subset \R\backslash\sigma(D_0)$.
 Our main result can be described as follows:
 If $(a,b)$ is a (non-trivial) spectral gap of
 the periodic operator $-\Delta + V$, acting in $L_2(\R^2)$, then
 for any compact interval  $[\alpha,\beta] \subset (a,b)$ with $\alpha <
 \beta$ there  is a $t \in (0,1)$ such that
 $\rho_{\text{surf}}(D_t,(\alpha,\beta))>0$. Upper bounds for the surface density of states are discussed
 in [HK].

 Our paper is organized as follows. Section 2 deals with
 dislocation on the real line. Here it is shown that the $k$-th gap
 in the essential spectrum of $H_t$ (if it is open) is crossed by
 effectively $k$ eigenvalues of $H_t$ as $t$ increases from $0$ to
 $1$. As an example, we discuss a periodic step potential in
 Section 3 where one can compute the eigenvalues of the dislocation
 operator numerically. Note that our calculations yield numbers
 which are exact up to finding the zeros of some transcendental
 functions. Related pictures can be found in [DPR] where a
 different numerical approach has been used.

 In Section 4 we adapt the method of Section 2 to the dislocation
 problem on the strip $\Sigma$. The results obtained for the strip then
 easily yield spectral information for the dislocation problem in
 the plane. Section 5 presents examples from the class of
 {\it muffin tin} potentials where one can ``see'' the motion of
 the eigenvalues rather directly for either translation in the
 $x$-direction or in the $y$-direction. Finally, we include an
 Appendix on regularity properties of the functions describing the
 eigenvalues of the dislocation operator $H_t$ in one
 dimension.

 For basic notation and definitions concerning self-adjoint operators
 in Hilbert space, we refer to [K, RS-I].\\[.25cm]\indent
 {\it Acknowledgements.} The authors would like to thank Andreas
 Ruschhaupt (Hannover) and Evgeni Korotyaev (St.\ Petersburg) for
 helpful discussions. The authors are particularly indebted to
 J\"urgen Voigt, Dresden, who kindly contributed Lemma A.6 in the
 Appendix.

\section{Dislocation on the real line}
In this section, we study perturbations of periodic Schr\"odinger
operators on the real line where the potential is obtained from a
periodic potential by a coordinate shift on the left half-axis.

Let $h_0$ denote the (unique) self-adjoint extension of
$-\frac{\d^2}{\dx^2}$ defined on $\Cci{\R}$. Our basic class of
potentials is given by
$$
\PP : = \left\{ V \in
L_{1,\text{loc}}(\R,\R);
\forall x \in \R :
 \, V(x+1) = V(x) \right\}.
 \eqno{(2.1)}
$$
Potentials $V \in \PP$ belong to the class
$L_{1,\text{loc,unif}}(\R)$ which coincides with the Kato-class on
the real line; in particular, any $V \in \P$ has relative
form-bound zero with respect to $h_0$ and thus the form-sum $H$ of
$h_0$ and $V \in \P$ is well defined (cf. [CFrKS]). For $V \in \P$
given, we define the dislocation potentials $W_t$ as in
eqn.~(1.1), for $0 \le t \le 1$;  as before, the form-sum $H_t$ of
$h_0$ and $W_t$ is well defined.

We begin with some well-known results pertaining to the spectrum
of $H = H_0$. As explained in [E, RS-IV], we have
$$
       \sigma(H) = \sigmaess(H) = \cup_{k=1}^\infty [\gamma_k,
       \gamma'_k],
  \eqno{(2.2)}
$$
where the $\gamma_k$ and $\gamma'_k$ satisfy $\gamma_k < \gamma_k'
\le \gamma_{k+1}$, for all $k \in \N$, and $\gamma_k \to \infty$
as $k\to\infty$. Moreover, the spectrum of $H$ is purely
absolutely continuous. The intervals $[\gamma_k,\gamma_k']$ are
called the {\it spectral bands} of $H$. The open intervals
$\Gamma_k := (\gamma_k',\gamma_{k+1})$ are the {\it spectral gaps}
of $H$; we say the $k$-th gap is {\it open} or {\it
non-degenerate} if $\gamma_{k+1} > \gamma_k'$.

In order to determine the essential spectrum of $H_t$ for $0<t<1$,
we introduce Dirichlet boundary conditions at $x=0$ for the
operator $H_0$ and at $x=0$ and $x = -t$ for $H_t$ to obtain the
operators
$$
           H_D=H^-\oplus H^+,\quad H_{t,D}=H_t^-\oplus H_{(-t,0)}\oplus H^+,
    \eqno{(2.3)}
$$
where $H^{\pm}$ acts in $\R^{\pm}$ with a Dirichlet boundary
condition at $0$, $H_t^-$ in $(-\infty,-t)$ with Dirichlet
boundary condition at $-t$ and $H_{(-t,0)}$ in $(-t,0)$ with
Dirichlet boundary conditions at $-t$ and $0$. Since $H_{(-t,0)}$
has purely discrete spectrum and since the operators $H_t^-$ and
$H^-$ are unitarily equivalent, we conclude that
$\sigmaess(H_D)=\sigmaess(H_{t,D})$. It is well known that
decoupling by (a finite number of) Dirichlet boundary conditions
leads to compact perturbations of the corresponding resolvents (in
fact, perturbations of finite rank) and thus  Weyl's essential
spectrum theorem yields $\sigmaess(H_D)=\sigmaess(H)$ and
$\sigmaess(H_{t,D})=\sigmaess(H_t)$.

In addition to the essential spectrum,  the operators $H_t$ may
have discrete eigenvalues below the infimum of the essential
spectrum and inside any (non-degenerate) gap, for $t \in (0,1)$;
these eigenvalues are simple. The eigenvalues of $H_t$ in the gaps
of $H$ depend continuously on $t$; cf.\ the Appendix for a brief
exposition of the relevant perturbational arguments, which are
fairly standard. A more complete and precise picture is
established in the following lemma which says that the discrete
eigenvalues of $H_t$ inside a given gap $\Gamma_k$ of $H$ can be
described by an (at most) countable, locally finite family of
continuous functions, defined on suitable subintervals of $[0,1]$.
\lem Let $k \in \N$ and suppose that the gap $\Gamma_k$ of $H$ is
open, i.e., $\gamma_k' < \gamma_{k+1}$. Then there is a (finite or
countable) family of continuous functions $f_j \colon
(\alpha_j,\beta_j) \to \Gamma_k$, where $0 \le \alpha_j < \beta_j
\le 1$, with the following properties:
\begin{enumerate}
\item[$(i)$] $f_j(t)$ is an eigenvalue of $H_t$, for all $\alpha_j
< t < \beta_j$ and for all $j$. Conversely, for any $t \in (0,1)$
and any eigenvalue $E \in \Gamma_k$ of $H_t$ there is a unique
index $j$ such that $f_j(t) = E$.
\item[$(ii)$] As $t \downarrow \alpha_j$ (or $t \uparrow
\beta_j$), the limit of $f_j(t)$ exists and belongs to the set
$\{\gamma_k',\gamma_{k+1}\}$.
\item[$(iii)$] For all but a finite number of indices $j$ the
range of $f_j$ does not intersect a given compact subinterval
$[a',b'] \subset \Gamma_k$.
\end{enumerate}
\endlem\rm
For the convenience of the reader, we include a proof in the
Appendix. Under stronger assumptions on $V$ one can show that the
eigenvalue branches are H\"older- or Lipschitz-continuous, or even
analytic; cf.\ the Appendix. Additional information on the
eigenvalue functions $f_j$ can be found in [K1, K2].

 It is our aim in this section to show that at least $k$ eigenvalues move
 from the upper to the lower edge of the $k$-th gap
 as the dislocation  parameter ranges from $0$ to $1$.
 Using the notation of Lemma 2.1
 and writing $f_i(\alpha_i) := \lim_{t\downarrow\alpha_i}f_i(t)$,
 $f_i(\beta_i) := \lim_{t\uparrow\beta_i}f_i(t)$, we now define
$$
  \NN_k := \#\{ i \mid f_i(\alpha_i) = \gamma_{k+1}, \,\,\, f_i(\beta_i) = \gamma_k' \}
           -  \#\{ i \mid f_i(\alpha_i) = \gamma_k', \,\,\, f_i(\beta_i) = \gamma_{k+1} \}.
  \eqno{(2.4)}
$$
Thus $\NN_k$ is precisely the number of eigenvalue branches of
$H_t$ that cross the $k$-th gap moving from the upper to the lower
edge minus the number crossing from the lower to the upper edge.
Put differently, $\NN_k$ is the spectral multiplicity which {\it
effectively} crosses the gap $\Gamma_k$  in downwards direction as
$t$ increases from $0$ to $1$.

Our main result in this section says that $\NN_k = k$, provided
the $k$-th gap is open:
\thm Let $V \in \PP$  and suppose that the $k$-th spectral gap of
$H$ is open, i.e., $\gamma_k' < \gamma_{k+1}$. Then $\NN_k = k$.
\endthm\rm
 Again, the results obtained by Korotyaev in [K1, K2] are more detailed; e.g.,
 it is shown that, for any $t \in (0,1)$, the dislocation operator $H_t$ has
 two unique states (an eigenvalue and a resonance) in any given gap of the
 periodic problem.
 On the other hand, our variational arguments are more flexible
 and allow an extension to higher dimensions, as will be seen in the sequel.
 In this sense, the importance of this section lies in testing our
 approach in the simplest possible case. For further reading concerning the spectral
 flow through the gaps of perturbed Schr\"odinger operators, we
 recommend [P, Saf].

The main idea of our proof---somewhat reminiscent of [DH,
ADH]---goes as follows: consider a sequence of approximations on
intervals $(-n-t,n)$ with associated operators $H_{n,t} =
-\frac{\text{d}^2}{\text{d}x^2}+ W_t$ with periodic boundary
conditions. We first observe that the gap $\Gamma_k$ is free of
eigenvalues of $H_{n,0}$ and $H_{n,1}$ since both operators are
obtained by restricting a periodic operator on the real line to
some interval of length equal to an entire multiple of the period,
with periodic boundary conditions. Second, the operators $H_{n,t}$
have purely discrete spectrum and it follows from Floquet theory
(cf. [E, RS-IV]) that $H_{n,0}$ has precisely $2n$ eigenvalues in
each band while $H_{n,1}$ has precisely $2n+1$ eigenvalues in each
band. As a consequence, effectively $k$ eigenvalues of $H_{n,t}$
must cross any fixed $E\in\Gamma_k$ as $t$ goes from $0$ to $1$.
To obtain the result of Theorem 2.2 we only have to take the limit
$n \to \infty$. Here we employ several technical lemmas. In the
first one, we show that the eigenvalues of the family $H_{n,t}$
depend continuously on the dislocation parameter.
\lem The eigenvalues of $H_{n,t}$ depend continuously on $t \in
[0,1]$.\endlem\rm
\proof We may assume that the eigenvalues of $H_{n,t}$ are
numbered according to min-max. Since the Hilbert space
$L_2(-n-t,n)$ depends on $t$, we use the unitary mappings
$$
  U_{n,t} \colon L_2(-n-t,n) \to L_2(-n,n),
  \qquad
  (U_{n,t} f)(x) := \sqrt{\sigma_{n,t}} f(\sigma_{n,t} x),
  \eqno{(2.5)}
$$
 where $\sigma_{n,t} :=  \frac{2n + t}{2n}$. Let ${\tilde H}_{n,t} := U_{n,t} H_{n,t}
U_{n,t}^{-1}$ and ${\tilde W}_t(x) := W_t(\sigma_{n,t} x)$ so that
(writing $\sigma = \sigma_{n,t}$)
$$
   {\tilde H}_{n,t} = \sigma^{-2} h_0 + {\tilde W}_t(x)
                  = \sigma^{-2}(h_0 + \sigma^2 {\tilde W}_t(x)).
   \eqno{(2.6)}
$$
It is easy to see that the mapping $[0,1] \ni t \mapsto \sigma^2
{\tilde W}_t \in L_1(-n,n)$ is continuous. Now the usual
perturbational and variational arguments for quadratic forms ([K]
and the Appendix) imply that the eigenvalues of $h_0 + \sigma^2
{\tilde W}_t$ depend continuously on $t$, and then the same is
true for the eigenvalues of $H_{n,t}$.
\endproof
The next lemma is to establish a connection between the spectra of
$H_t$ and $H_{n,t}$ for $0 \le t \le 1$ and $n$ large.  In the
proof and henceforth, we will make use of the following cut-off
functions: We pick some $\phi\in\Cci{-2,2}$ with $0\leq\phi\leq 1$
and $\phi(x)=1$ for $|x| \le 1$. For $k \in (0,\infty)$ we then
define $\phi_k(x):=\phi(x/k)$ so that
$\supp\phi_k\subset(-2k,2k)$, $\phi_k(x)=1$ for $|x|\leq k$,
$|\phi_k'(x)|\leq Ck^{-1}$ and $|\phi_k''(x)|\leq Ck^{-2}$.
Finally, we let $\psi_k:=1-\phi_k$. For any self-adjoint operator
$T$ we denote the spectral projection associated with an interval
$I \subset \R$ by $P_I(T)$ and we write $\dim\, P_I(T)$ to denote
the dimension of the range of the projection $P_I(T)$.
\lem Let $k \in \N$ with $\Gamma_k\neq\emptyset$.
 Let $t \in (0,1)$ and suppose that $\eta_1 < \eta_2 \in \Gamma_k$
are such that $\eta_1, \eta_2 \notin \sigma(H_t)$. Then there is
an $n_0 \in \N$ such that $\eta_1, \eta_2 \notin \sigma(H_{n,t})$
for $n \ge n_0$, and\rm
$$
    \dim\, P_{(\eta_1, \eta_2)}(H_t) = \dim\, P_{(\eta_1, \eta_2)}(H_{n,t}),
  \qquad n \ge n_0.
   \eqno{(2.7)}
$$
\endlem\rm
\proof In the subsequent calculations, we always take $k := n/4$,
for $n\in\N$.

(1) Let $E \in (\eta_1,\eta_2) \cap \sigma(H_t)$ with associated
normalized eigenfunction $u$.  Then $u_k:=\phi_ku\in D(H_{n,t})$,
$H_{n,t}u_k=H_tu_k$ and $\norm{u_k}\to 1$ as $n \to \infty$. Since
$$
\norm{H_{n,t}u_k-E u_k}\leq 2\cdot\infnorm{\phi_k'}\norm{u'}
+\infnorm{\phi_k''}\norm{u},\eqno{(2.8)}
$$
it is now easy to conclude that $ \dim\, P_{(\eta_1,
\eta_2)}(H_{n,t}) \ge  \dim\, P_{(\eta_1, \eta_2)}(H_t)$ for $n$
large.

(2) We next assume for a contradiction that $\eta \in \Gamma_k$
satisfies
 $\eta \in \sigma(H_{n,t})$ for infinitely many $n \in \N$.
Then there is a subsequence $(n_j)_{j\in\N} \subset \N$ s.th.\
$\eta \in \sigma(H_{n_j,t})$; we let $u_{n_j,t} \in D(H_{n_j,t})$
denote a normalized eigenfunction and set
$$
   v_{1,n_j} := \phi_{k_j} u_{n_j,t},
    \qquad    v_{2,n_j} := \psi_{k_j}  u_{n_j,t},\eqno{(2.9)}
$$
so that $v_{1,n_j} \in D(H_t)$ and $\norm{(H_t - \eta)v_{1,n_j}}
\to 0$ as $j \to \infty$ by a similar estimate as in part (1) (and
using a simple bound for $\norm{u_{n,t}'}$ which follows from the
fact that $V$ has relative form-bound zero w.r.t.\ $h_0$.) Let us
now show that $v_{2,n_j} \to 0$ (and hence $\norm{v_{1,n_j}} \to
1$) as $j \to \infty$:
 The function
$$
\tilde{v}_{2,n_j}:= \left\{%
\begin{array}{lll}
  v_{2,n_j}(x),   &&  x \geq 0, \\
  v_{2,n_j}(x-t), && x < 0, \\
\end{array}%
\right.
  \eqno{(2.10)}
$$
 belongs to the domain of $H_{n_j,0}$ and
 $H_{n_j,0}\tilde{v}_{2,n_j}=[H_{n_j,t}v_{2,n_j}]^\sim$\,, where
 $[\cdot]^\sim$ is defined in analogy with eqn.~(2.10). Since we also have
 $(H_{n_j,t} - \eta)v_{2,n_j} \to 0$, as $j \to \infty$,
we see  that  $(H_{n_j,0}-\eta)\tilde v_{2,n_j} \to 0$. But
$\dist(\eta,\sigma(H_{n,0})) \ge \delta_0 > 0$ for all $n$ and the
Spectral
 Theorem implies that $\norm{\tilde{v}_{2,n_j}}\to 0$ as
 $j\to\infty$. We have thus shown that  $\norm{v_{1,n_j}} \to 1$ and
 $\norm{(H_t - \eta)v_{1,n_j}} \to 0$ which implies that $\eta \in
 \sigma(H_t)$.

(3) It remains to show that $ \dim\, P_{(\eta_1, \eta_2)}(H_{n,t})
\le  \dim\, P_{(\eta_1, \eta_2)}(H_t)$, for $n$ large. The proof
by contradiction follows the lines of part (2); instead of a
sequence of functions $u_{n_j}$ we work with an orthonormal system
$u_{n_j}^{(1)}, \ldots, u_{n_j}^{(\ell)}$ of eigenfunctions where
$\ell = \dim\, P_{(\eta_1,\eta_2)}(H_t + 1)$. We leave the details
to the reader.
\endproof
\rem In fact, using standard exponential decay estimates for
resolvents of Schr\"odinger operators, cf.~[S], it can be shown
that the eigenvalues of $H_t$ and $H_{n,t}$ in the gap $\Gamma_k$
are exponentially close, for $n$ large; e.g., if $E \in
\sigma(H_t) \cap \Gamma_k$ for some $t \in (0,1)$, then there are
constants $c \ge 0$ and $\alpha > 0$ s.th.\ the operators
$H_{n,t}$ have an eigenvalue in $(E- c \ee^{-\alpha n}, E + c
\ee^{-\alpha n})$, for $n$ large. There is a similar converse
statement with the roles of $H_t$ and $H_{n,t}$ exchanged; cf.\
also Remark 4.2 for further discussion.\endrem\rm

The desired connection between the spectral flow for $(H_{n,t})_{0
\le t \le 1}$ and $(H_t)_{0 \le t \le 1}$ is obtained by applying
Lemma 2.4 at suitable $t_i \in [0,1]$ and $\eta_{1,i} < \eta_{2,i}
\in \Gamma_k$. We now construct an appropriate partition of the
parameter interval $[0,1]$.
\lem Let $k \in \N$ with $\Gamma_k\neq\emptyset$.
  Then there exists a partition $0 = t_0
< t_1 < \ldots < t_{K-1} < t_K = 1$ and there exist $E_j \in
\Gamma_k$ and $n_0 \in \N$ such that
$$
  E_j \notin \sigma(H_t) \cup \sigma(H_{n,t}), \qquad \forall t \in [t_{j-1},
  t_j], \quad j=1, \ldots, K, \quad n \ge n_0.\eqno{(2.11)}
$$
\endlem\rm
\proof For any $t \in [0,1]$ there exists $\eta_t \in \Gamma_k$
such that $\eta_t \notin \sigma(H_t)$. Since the spectrum of $H_t$
depends continuously on the parameter $t$ there also exists $\eps
= \eps_t > 0$ such that $\eta_t \notin \sigma(H_\tau)$ for all
$\tau \in (t -\eps_t, t + \eps_t)$. By compactness, we can find a
partition $(\tau_j)_{0 \le j \le K}$ (with $\tau_0 = 0$, $\tau_K =
1$) such that the intervals $(\tau_j - \eps_j, \tau_j + \eps_j)$
cover $[0,1]$. Set $E_j := \eta_{\tau_j}$. We next pick arbitrary
points $t_j \in (\tau_j , \tau_j + \eps_j) \cap    (\tau_{j+1}
-\eps_{j+1}, \tau_{j+1})$, for $j=1, \ldots, K-1$, set $t_0=0$,
$t_K=1$ and see that $E_j \notin \sigma(H_t)$ for $t_{j-1} \le t
\le t_j$, $j = 1, \ldots, K$.
 By Lemma 2.4, using Lemma 2.3 combined with a simple compactness argument,
 we then find that we also have $E_j \notin \sigma(H_{n,t})$
for  $t \in [t_{j-1},t_j]$ and $n$ large.
\endproof
We are now ready for the proof of Theorem 2.2.\\[.2cm]
\emph{Proof of Theorem 2.2.} Let $E_j$ be as in Lemma 2.6 and $\NN_k$
as in
 eqn.~(2.4). We fix some ${\tilde E} \in \Gamma_k$
such that ${\tilde E} > E_j$ for $j = 0,\ldots, K$ and ${\tilde E}
\notin \sigma(H_{t_j}) \cup \sigma(H_{n,t_j})$ for $j = 0, \ldots,
K$ and for all $n$ large. It is then easy to see that
$$
  \NN_k = \sum_{j=1}^K \left( \dim\, P_{(E_j, {\tilde E})}(H_{t_j})
             -  \dim\, P_{(E_j, {\tilde E})}(H_{t_{j-1}}) \right)\eqno{(2.12)}
$$
and that
\begin{align}
  \dim\, P_{(-\infty, {\tilde E})}(H_{n,1}) & -  \dim\, P_{(-\infty, {\tilde
   E})}(H_{n,0})  \nonumber\\
 &   =  \sum_{j=1}^K \bigg( \dim\, P_{(E_j, {\tilde E})}(H_{n,t_j})
             -  \dim\, P_{(E_j, {\tilde E})}(H_{n,t_{j-1}})
\bigg) .\nonumber
\end{align}
\vspace{-.8cm}
$$\eqno{(2.13)}$$
The LHS of (2.13) equation equals $k$. Furthermore, by Lemma
2.4, we have
$$
  \dim\, P_{(E_j, {\tilde E})}(H_{t_j}) =  \dim\, P_{(E_j, {\tilde E})}(H_{n,t_j})\eqno{(2.14)}
$$
for all $j$ and all $n$ large, and the desired result follows.
\hfill$\square$
\section{A one-dimensional periodic step
potential}
In this section, we study the one-dimensional $2\pi$-periodic
potential
$$
V(x):=\left\{%
\begin{array}{lll}
  -1, && x\in[0,\pi], \\
  1,  && x\in(\pi,2\pi). \\
\end{array}%
\right.\eqno{(3.1)}
$$
(While the other sections of this paper deal with $1$-periodic potentials,
we have preferred to work here with period $2\pi$ in order to keep
the explicit calculations done by hand as simple as possible.)
To obtain the band-gap structure of $H=-\frac{\d^2}{\dx^2}+V$, we
compute the discriminant function
$$
D(E):=\phi_1(2\pi;E)+\phi_2'(2\pi;E)=\hbox{tr}
\left(%
\begin{array}{cc}
  \phi_1(2\pi;E) & \phi_1'(2\pi;E) \\
  \phi_2(2\pi;E) & \phi_2'(2\pi;E) \\
\end{array}%
\right)\eqno{(3.2)}
$$
where $\phi_1(\,\cdot\,;E)$ and $\phi_2(\,\cdot\,;E)$ solve the
equation
$$-u''+(V-E)u=0\eqno{(3.3)}$$
and satisfy the boundary conditions
$$\phi_1(0;E)=\phi_2'(0;E)=1\quad\hbox{and}\quad\phi_1'(0;E)=\phi_2(0;E)=0.\eqno{(3.4)}$$
The matrix $M(E)$ on the RHS of (3.2) is called the {\it monodromy
matrix}. A simple computation shows that $[-1/2,1/2]\subset\Gamma_1$,
where $\Gamma_1$ is the first spectral gap of $H$ (with numbering
according to Floquet theory). Note that the gap edges of
$\Gamma_1$ also equal the first eigenvalue in the (semi-)periodic
eigenvalue problem for $-\frac{\text{d}^2}{\text{d}x^2}+V$ in
$L_2(0,2\pi)$, cf., e.g., [E, CL].

As explained in [E, RS-IV], for any $E\notin\sigma(H)$, there are
two solutions $\phi_\pm(x;E)\in C^1(\R)$, square integrable at
$\pm\infty$, of (3.3); in fact, the functions $\phi_\pm(x;E)$ are
exponentially decaying at $\pm\infty$ and exponentially increasing
at $\mp\infty$. In our example, the dislocation potential $W_t$
for $t\in(0,1)$ will produce a bound state at $E$ if and only if
the boundary conditions coming from $\phi_+(0;E)$ and
$\phi_-(t;E)$ match up, i.e.,
$$\phi_-(t;E)=\phi_+(0;E)\quad\text{and}\quad\phi_-'(t;E)=\phi_+'(0;E).\eqno{(3.5)}$$
An equivalent condition for (3.5) is the equality of the ratio
functions $\frac{\phi_-(t;E)}{\phi_-'(t;E)}$ and
$\frac{\phi_+(0;E)}{\phi_+'(0;E)}$, cf. [DPR].
We compute the Floquet solutions $\varphi_{\pm}$ by solving the
equation $-u''+(V-E)u=0$ for $x<0$ and $x>0$ and for $E$ varying
in $[-1/2,1/2]$, assuming that $(u(0),u'(0))$ equals an
appropriate eigenvector of $M(E)$. Note that, since $D(E)<-2$,
both eigenvalues of $M(E)$ are negative and not equal to $-1$.
Finally, we divide $[-1/2,1/2]$ into 100 subintervals of equal
length and compute numerical values for $t$ such that
$$\left|\frac{\phi_-(t;E)}{\phi_-'(t;E)}-\frac{\phi_+(0;E)}{\phi_+'(0;E)}\right|<\eps,\eqno{(3.6)}$$
where the error $\eps>0$ is small. This leads to the following
plot of $t\mapsto E(t)$, see Fig.~1.
\begin{figure}[H]
\begin{center}
\includegraphics[width=8cm]{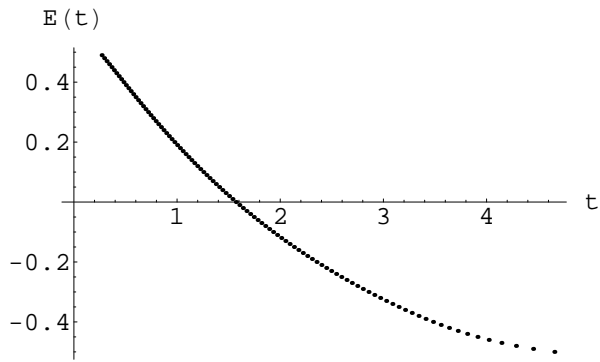}
\caption{An eigenvalue branch of $H_t$ in the first spectral gap.}
\end{center}
\end{figure}
\section{Periodic potentials on the strip
and the plane}
Let $V \colon \R^2 \to \R$ be $\Z^2$-periodic and Lipschitz-continuous and let $\Sigma = \R \times (0,1)$ denote the infinite
strip of width $1$. We denote by $S_t$ the (self-adjoint) operator
$-\Delta + W_t$, acting in $L_2(\Sigma)$, with periodic boundary
conditions in the $y$-variable and with $W_t$ defined as in
eqn.~(1.2);
 again, the parameter
$t$ ranges between $0$ and $1$. Since $S_0$ is periodic in the
$x$-variable, its spectrum has a band-gap structure.

We first observe that the essential spectrum of the family $S_t$
does not depend on the parameter $t$, i.e.,
$\sigmaess(S_t)=\sigmaess(S_0)$ for all $t\in[0,1]$. As in Section
2, this follows from the compactness of
$(S_t-c)^{-1}-(S_{t,D}-c)^{-1}$, where $S_{t,D}$ is $S_t$ with an
additional Dirichlet boundary condition at $x=0$, say. (While, in
one dimension, adding in a Dirichlet boundary condition at a
single point causes a rank-one perturbation of the resolvent, the
resolvent difference is now Hilbert-Schmidt, which can be seen
from the following well-known line of argument: If
$-\Delta_\Sigma$ denotes the (negative) Laplacian in $L_2(\Sigma)$
and $-\Delta_{\Sigma;D}$ is the (negative) Laplacian in
$L_2(\Sigma)$ with an additional Dirichlet boundary condition at
$x=0$, then $(-\Delta_\Sigma + 1)^{-1}
  - (-\Delta_{\Sigma;D} + 1)^{-1}$ has an integral kernel which can be
written down explicitly using the Green's function for
$-\Delta_\Sigma$ and the reflection principle.)

While the essential spectrum of the family $S_t$ does not change
as $t$ ranges through $[0,1]$, $S_t$ will have discrete
eigenvalues in the spectral gaps of $S_0$ for appropriate values
of $t$.
We have the following result.
\thm Let $(a,b)$, $a<b$, denote a spectral gap of $S_t$ and let $E
\in (a,b)$. Then there exists $t = t_E \in (0,1)$ such that $E$ is
a discrete eigenvalue of $S_t$.
\endthm\rm
\proof (1) As on the real line, we work with approximating
problems on finite size sections of the infinite strip $\Sigma$.
Let
$$
\Sigma_{n,t}:=(-n-t,n)\times(0,1),\quad n\in\N,\eqno{(4.1)}
$$
and consider $S_{n,t}:=-\Delta+W_t$ acting in $L_2(\Sigma_{n,t})$
with periodic boundary conditions in both coordinates. The
operator $S_{n,t}$ has compact resolvent and purely discrete
spectrum accumulating only at $+\infty$. The rectangles
$\Sigma_{n,0}$ (respectively, $\Sigma_{n,1}$) consist of $2n$
(respectively, $2n+1$) period cells. By routine arguments (see,
e.g., [RS-IV, E]), the number of eigenvalues below the gap $(a,b)$
is an integer multiple of the number of cells in these rectangles;
we conclude, that eigenvalues of $S_{n,t}$ must cross the gap as
$t$ increases from $0$ to $1$.\\
(2) Let $E\in(a,b)$. According to (1), for any $n\in\N$ we can
find $t_n\in(0,1)$ such that $E\in\sigmadisc(S_{n,t_n})$; then
there are eigenfunctions $u_n\in D(S_{n,t_n})$ with $S_{n,t_n}u_n
= Eu_n$, $\norm{u_n}=1$, and $\norm{\nabla u_n} \le C$ for some
constant $C \ge 0$. We now choose cut-off functions $\phi_n$ as in
Section 2 and denote the natural extension to $\R^2$ again by
$\phi_n$. We also let $\psi_n = 1 - \phi_n$. Clearly,
$$
   \norm{(S_{t_n} - E)(\phi_{n/4} u_n)}, \hskip1ex
   \norm{(S_{n,t_n} - E)(\psi_{n/4} u_n)} \le c/n,
   \eqno{(4.2)}
$$
for some $c \ge 0$. There is a subsequence $(t_{n_j})_{j\in\N}
\subset (t_n)_{n\in\N}$ and $\cl t \in [0,1]$ s.th.\ $t_{n_j} \to
\cl t$ as $j\to\infty$. Since $V$ is Lipschitz, we may infer from
$(4.2)$ that
$$
   \norm{(S_{\cl t} - E)(\phi_{n_j/4} u_{n_j})} \to 0, \qquad j \to \infty,
   \eqno{(4.3)}
$$
and it remains to show that $\norm{\psi_{n/4} u_n} \to 0$ so that
$\norm{\phi_{n/4} u_n} \to 1$. We associate with functions $v
\colon \Sigma_{n,t} \to \C$ functions $\tilde{v} \colon
\Sigma_{n,0} \to \C$ by
$$
   \tilde{v}(x,y) :=
   \left\{%
\begin{array}{lll}
  v(x,y),   && x>0, \\
  v(x-t,y), && x < 0,\\
\end{array}%
\right.
  \eqno{(4.4)}
$$
in analogy with eqn.~(2.10). Then $[\psi_{n/4} u_n]^{\sim} \in
D(S_{n,0})$ and
$$
   \norm{(S_{n,0} - E)[\psi_{n/4} u_n]^{\sim}}
             = \norm{(S_{n,t_n} - E)(\psi_{n/4} u_n)} \le c/n.
 \eqno{(4.5)}
$$
Since $(a,b) \cap \sigma(S_{n,0}) = \emptyset$ for all $n\in\N$,
and since $E \in (a,b)$, the Spectral Theorem implies that
$[\psi_{n/4} u_n]^{\sim} \to 0$ (and therefore also $\psi_{n/4}u_n
\to 0$) as $n \to \infty$.

We therefore have shown that the functions $v_{n_j} :=
\phi_{n_j/4} u_{n_j}$ for $j\in\N$ satisfy $\norm{(S_{\cl t} - E)
v_{n_j}} \to 0$ and $\norm{v_{n_j}} \to 1$ as $j \to \infty$ which
implies $E \in \sigma(S_{\cl t})$.
\endproof
\rem By a well-known line of argument, one can obtain {\it
exponential localization} of the eigenfunctions of $S_t$ near the
interface $\{(x,y) \mid x = 0 \}$. Since we will use exponential
localization in a more systematic way in the forthcoming paper
[HK] we only give a brief sketch here: Suppose that $E \in (a,b)$
and $t \in (0,1)$ satisfy $E \in \sigma(S_t)$. Let $u \in D(S_0) =
D(S_t)$ denote a normalized eigenfunction and let $\phi_n$, $n \in
\N$, be as in the proof of Theorem 4.1. As above, we have
$$
    (S_t - E) (\phi_n u) = -2 \nabla \phi_n \cdot \nabla u - (\Delta \phi_n) u
    = : r_n,
 \eqno{(4.6)}
$$
where $\norm{r_n} \le c/n$, for $n \in \N$. Since $r_n$ has
support in the interval $(-2n-1, 2n)$ we now see that there exist
constants $C \ge 0$ and $\alpha > 0$ such that
$$
   \norm{\chi_{|x| \ge 4n} u}
    \le \norm{\chi_{|x| \ge 4n} (S_t - E)^{-1} r_n}
    \le C \ee^{-\alpha n},
 \eqno{(4.7)}
$$
by standard exponential decay estimates for the resolvent kernel
of Schr\"odinger operators (cf., e.g., [S], [HK]).
\endrem\rm
We now turn to the dislocation problem on the plane $\R^2$ where
we study the operators
$$
           D_t=-\Delta+W_t,\quad 0\leq t\leq 1.
           \eqno{(4.8)}
$$
Denote by $S_t(\theta)$ the operator $S_t$ with $\theta$-periodic
boundary conditions in the $y$-variable. Since $W_t$ is periodic
with respect to $y$, we have
$$
D_t\simeq\int_{[0,2\pi]}^{\oplus}S_t(\theta)\frac{\d\theta}{2\pi},
\eqno{(4.9)}
$$
and hence the spectrum of $D_t$ has a band-gap structure;
furthermore, $D_t$ has no singular continuous part, cf.\ [DS, FS].
As for the spectrum of $S_t$ inside the gaps of $S_0$, Theorem 4.1
leads to the following result.
\thm Let $(a,b)$ denote a spectral gap of $D_0$, $a > \inf
\sigmaess(D_0)$, and let $E \in (a,b)$. Then there exists $t = t_E
\in (0,1)$ with $E \in \sigma(D_t)$.
\endthm\rm
\proof Let $\phi_n u_n\in D(S_t)$ as in part (2) of the proof of
Theorem 4.1 denote an approximate solution of the eigenvalue
problem for $S_t$ and $E$. We extend $u_n$ to a function ${\tilde
u}_n(x,y)$ on $\R^2$ which is periodic in $y$. Writing $\Phi_n =
\Phi_n(x,y) := \phi_n(x) \phi_n(y)$ we compute
\begin{align}
(D_t - E) (\Phi_n\tilde u_n)
   & = \left(- \partial_x^2 - \partial_y^2 + W_t - E \right) (\phi_n(x)
   \phi_n(y) \tilde u_n(x,y)) \nonumber\\
   & = \phi_n(y) {[(S_t - E) (\phi_n(x) u_n)]^{\sim} }
       - \phi_n(x) \left(2 \phi_n'(y) \partial_y {\tilde u}_n + \phi_n''(y) {\tilde u}_n
         \right). \nonumber
\end{align}
\vspace{-.65cm}
$$\eqno{(4.10)}$$
The norms of the three terms on the RHS can be estimated (up to  a
constant which is independent of $n$) by $\eps n$, ${1 \over n} n$
and ${1 \over n^2} n$, respectively, and we see that
$$
  \norm{(D_t - E) (\Phi_n\tilde u_n)} \le c_0 (1 + n\eps),
  \eqno{(4.11)}
$$
while $\norm{\Phi_n\tilde u_n} \ge c_0 n$ with a constant $c_0 >
0$. This implies the desired result.
\endproof
\rem We learn from the above proof that there are functions
$$
   v_n = v_n(x,y) := {1 \over \norm{\Phi_n\tilde u_n}} \Phi_n\tilde u_n
   \eqno{(4.12)}
$$
that satisfy $\norm{v_n} =1$, $\supp v_n \subset [-n,n]^2$ and
$$
   (D_t - E) v_n \to 0, \qquad n \to \infty.
  \eqno{(4.13)}
$$
These functions play a key role in our analysis of the rotation
problem at small angle in [HK].
\endrem\rm
We finally turn to a brief discussion of the i.d.s.\ (the
integrated density of states [V]) for the dislocation operators $D_t$.
We adopt the natural distinction of [DS, EKSchrS, KS] between {\it
bulk} and {\it surface} states. Roughly speaking, the bulk states
correspond to states away from the interface with energies in the
spectral bands while the surface states for $0< t < 1$ are
produced by the interface and are (exponentially) localized near
the interface.
 The (integrated) density of states measures for the bulk and surface states
 use a different scaling factor in the following
definition: restricting $D_t$ to large squares $Q_n = (-n,n)^2$
and taking Dirichlet boundary conditions, we obtain the operators
$D^{(n)}_t$. For $I \subset \R$ an open interval, let
$N(I,D^{(n)}_t)$ denote the number of eigenvalues of $D^{(n)}_t$
in $I$, counting multiplicities. We then define for open intervals
$I \subset \R$ and $J \subset \R \setminus \sigma (D_0)$ with
$\overline{J} \subset  \R \setminus \sigma (D_0)$
$$
   \rho_{\text{bulk}}(I, D_t) = \lim_{n \to \infty} \frac{1}{4n^2} N(I,D^{(n)}_t),
   \quad
     \rho_{\text{surf}}(J, D_t) = \lim_{n \to \infty} \frac{1}{2n}
   N(J,D^{(n)}_t).
 \eqno{(4.14)}
  $$
The existence of the limits in (4.14) has been established in
[EKSchrS, KS] for ergodic Schr\"odinger operators. Note that the
surface density of states measure is defined (and possibly
non-zero) for subintervals of the spectral bands, but then
eqn.~(4.14) is not suited to capture the surface states (cf.\
[EKSchrS, KS]).

The fact that the surface density of states exists does not mean
it is non-zero and there are only rare examples where we know
$\rho_{\text{surf}}$ to be non-trivial. It is one of the main
results of the present paper to show that dislocation moves enough
states through the gap to have a non-trivial surface density of
states, for suitable parameters $t$. Indeed, it is now
easy to derive the following result:
\cor Let $(a,b)$ be a spectral gap of $D_0$ with $a > \inf
\sigmaess(D_0)$,  and let $\emptyset \ne J \subset (a,b)$ be an
open interval. Then there is a $t \in (0,1)$ such that
$\rho_{\text{\rm surf}}(J,D_t) > 0$.
\endcor\rm
\proof
Let $[\alpha,\beta] \subset J$ with $\alpha < \beta$, fix
$E \in (\alpha,\beta)$, and let $0 < \eps < \min \{ E-\alpha,
\beta - E\}$. By Theorem~4.3 and Remark~4.4 there exist $t = t_E
\in (0,1)$ and a function $u_0$ in the domain of $D_t$ satisfying
$\norm{u_0} = 1$, $\supp u_0$ compact, and $\norm{(D_t - E) u_0} < \eps$.
Let $\nu \in \N$ be such that $\supp u_0 \subset (-\nu,\nu)^2$; note that,
in the present proof, $\nu$ corresponds to the $n$ of Remark 4.4.
We then see that the functions $\varphi_k$, defined by
$\varphi_k(x,y) := u_0(x, y - 2 k \nu)$ for $k \in \N$, have pairwise
disjoint supports, are of norm $1$, and satisfy $\norm{(D_t - E) \phi_k} <
\eps$. Furthermore, we have $\supp \varphi_k \subset (-n,n)^2$ provided
$(2k + 1) \nu < n$. Denoting
 $\MM_n := \hbox{\rm span} \{ \phi_k \mid k \in \N, \> k \le {1 \over
   2}  ({n \over \nu} - 1) \}$,
 it is clear that $\dim \MM_n \ge  n/(3\nu)$, for all $n$ large.
 Let $\NN_n$ denote the range of the spectral projection
$P_{(\alpha,\beta)}(D_t^{(n)})$ of $D_t^{(n)}$ associated with the interval
$(\alpha,\beta)$; we will show that $\dim \NN_n \ge \dim \MM_n$ which
implies the desired result. If we assume for a contradiction
 that $\dim \NN_n < \dim \MM_n$ for some $n \in \N$,
 we can find a function $v \in \MM_n \cap \NN_n^\perp$ of norm $1$. By
the Spectral Theorem, $|\!|(D_t^{(n)} - E) v |\!| \ge \eps$. On the
other hand, $v$ is a finite linear combination of the $\phi_k$,
which implies $|\!|(D_t^{(n)} - E) v|\!| < \eps$.
\endproof
We will continue the discussion of bulk versus surface states in
the companion paper [HK] where a corresponding upper bound of the
form $N(J, D_t^{(n)}) \le c n \log n$ is provided.
\section{Muffin tin potentials}
Here we present some simple examples where one can see the
behavior of surface states directly. We will deal with
$\Z^2$-periodic muffin tin potentials of infinite height (or
depth) on the plane $\R^2$ which can be specified by fixing a
radius $0 < r < 1/2$ for the discs where the potential vanishes,
and the center $P_0 = (x_0, y_0) \in [0,1)^2$ for the generic
disc. In other words, we consider the periodic sets
$$
   \Omega_{r,P_0} := \cup_{(i,j) \in \Z^2} B_r(P_0 + (i,j)),
$$
and we let $V = V_{r,P_0}$ be zero on $\Omega_{r,P_0}$ while we
assume that $V$ is infinite on $\R^2 \setminus \Omega_{r,P_0}$. If
$H_{ij}$ is the Dirichlet Laplacian of the disc $B_r(P_0 +
(i,j))$, then the form-sum of $-\Delta$ and $V_{r,P_0}$ is
$\oplus_{(i,j)\in \Z^2} H_{ij}$. Without loss of generality, we
may assume $y_0 = 0$ henceforth.\\[.25cm]
{\bf (1) Dislocation in the $x$-direction.} Here muffin tin
potentials yield an illustration for some of the phenomena
encountered in Section 4. In the simplest case we would take $x_0
= 1/2$ so that the disks $B_r(1/2 + i, j)$, for $i \in \N_0$ and
$j \in \Z$, will not intersect or touch the interface $\{(x,y)
\mid x=0\}$. Defining the dislocation potential $W_t$ as in
Section 4, we see that there are bulk states given by the
Dirichlet eigenvalues of all the discs that do not meet the
interface, and there may be surface states given as the Dirichlet
eigenvalues of the sets $B_r(1/2 - t, j) \cap \{ x < 0\}$ for $j
\in \Z$ and $1/2 - r < t < 1/2 + r $.

More precisely, let $\mu_k = \mu_k(r)$ denote the Dirichlet
eigenvalues of the Laplacian on the disc of radius $r$, ordered by
min-max and repeated according to their respective multiplicities.
The Dirichlet eigenvalues of the domains $B_r(1/2 - t, 0) \cap \{x
< 0\}$, $1/2 - r < t < 1/2 + r$, are denoted as $\lambda_k(t) =
\lambda_k(t,r)$; they are continuous, monotonically decreasing
functions of $t$ and converge to $\mu_k$ as $t \uparrow 1/2 + r$
and to $+\infty$ as $t \downarrow 1/2 - r$. In this simple model,
the eigenvalues $\mu_k$ correspond to the bands of a periodic
operator. We see that the gaps are crossed by surface states as
$t$ increases from $0$ to $1$, in accordance with the results of
Section 4 (Corollary 4.5).

Along the same lines, one can easily analyze examples where $x_0$
is different from $1/2$; here more complicated geometric shapes
may come into play. In [HK] we will again use muffin tin
potentials as examples for the rotation problem. In that paper, we
will also discuss approximations by muffin tin potentials of
height $n$ and their limit as $n \to \infty$.\\[.25cm]
{\bf (2) Dislocation in the $y$-direction.} This problem has not
been considered so far. We include a brief discussion of this case
for two reasons: on the one side, we observe a new phenomenon
which did not appear so far; on the other hand, one can see from
our example that, presumably, there is no general theorem for
translation of the left half-plane in the $y$-direction.

 Let $V = V_r$ denote the muffin tin potential defined above, with $x_0 = y_0 = 0$.
 We then let ${\tilde W}_t$ coincide with
 $V$ in the right half-plane, while we take ${\tilde W}_t(x,y)
 = V(x, y - t)$ in the left half-plane. At the interface $\{x=0\}$ we see
 half-discs on the left and on the right with the half-discs on
 the right being fixed while the half-discs on the left are shifted
 by $t$ in the $y$-direction. The surface states correspond to the
 states of the Dirichlet Laplacian on the union
 $\Omega_{t,r;\text{surf}}$
 of these half-discs. There are two cases: either $\Omega_{t,r;\text{surf}}$
 is connected and  we have a scattering channel along the interface,
 or $\Omega_{t,r;\text{surf}}$ is  the disjoint union of a sequence of
 bounded domains; cf.\ Figure 2. In the second case, the eigenvalues on such
 domains start at the Dirichlet eigenvalues of the disc of radius $r$,
 increase up to the corresponding eigenvalues of a half-disc, and then
 move down again to where they started. For $1/4 < r < 1/2$, the picture
 is more complicated: If we let  $\tau_0 = 1 - 2r$, $\tau_1 = 2r$, we find
 that the sets $\Omega_{t,r;\text{surf}}$ are disconnected for
 $0\le t \le \tau_0$ and for $\tau_1 \le t \le 1$; for $\tau_0 < t < \tau_1$,
 however, $\Omega_{t,r;\text{surf}}$ is connected and forms a
 periodic wave guide with purely a.c.\ spectrum [SW]; cf.\ also [DS].
 We therefore observe a dramatic change in the spectrum
 of the dislocation operators: for $t \in [0,\tau_0] \cup [\tau_1,1]$ the surface states
 in the gap are given by eigenvalues of infinite multiplicity
 while for  $t \in (\tau_0, \tau_1)$ the surface states form bands of a.c.\ spectrum in the gaps.

\begin{figure}[H]
\begin{center}
\includegraphics[width=5.2cm]{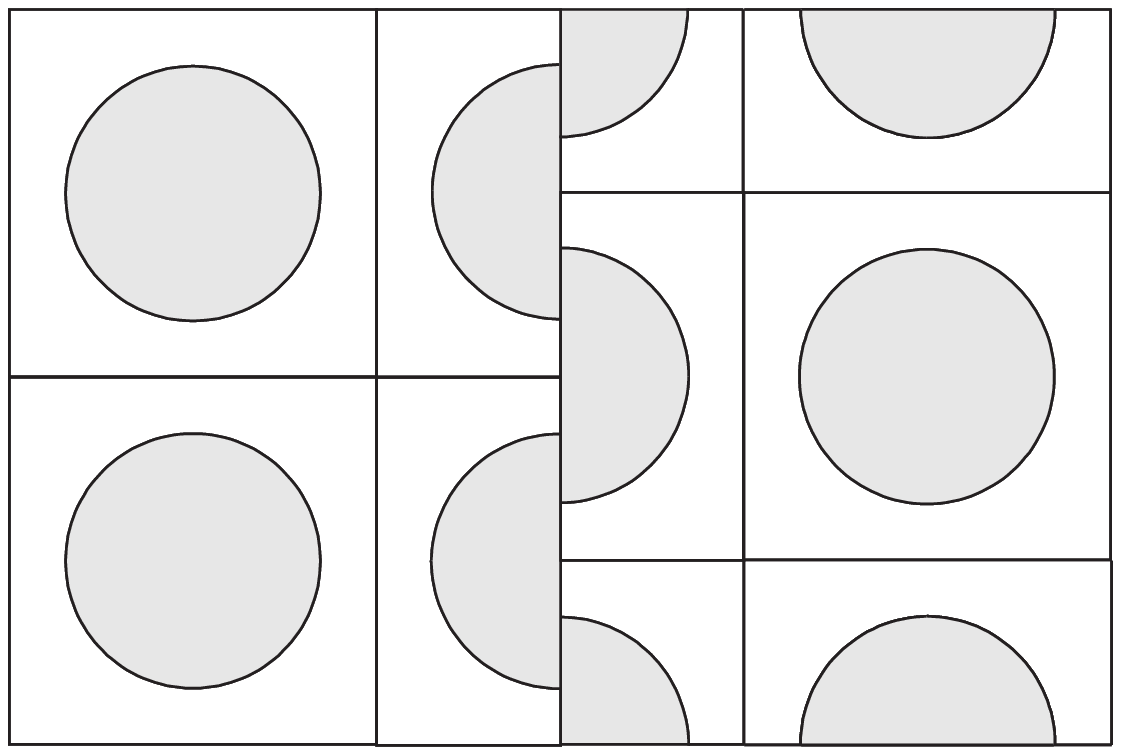}
\includegraphics[width=5.2cm]{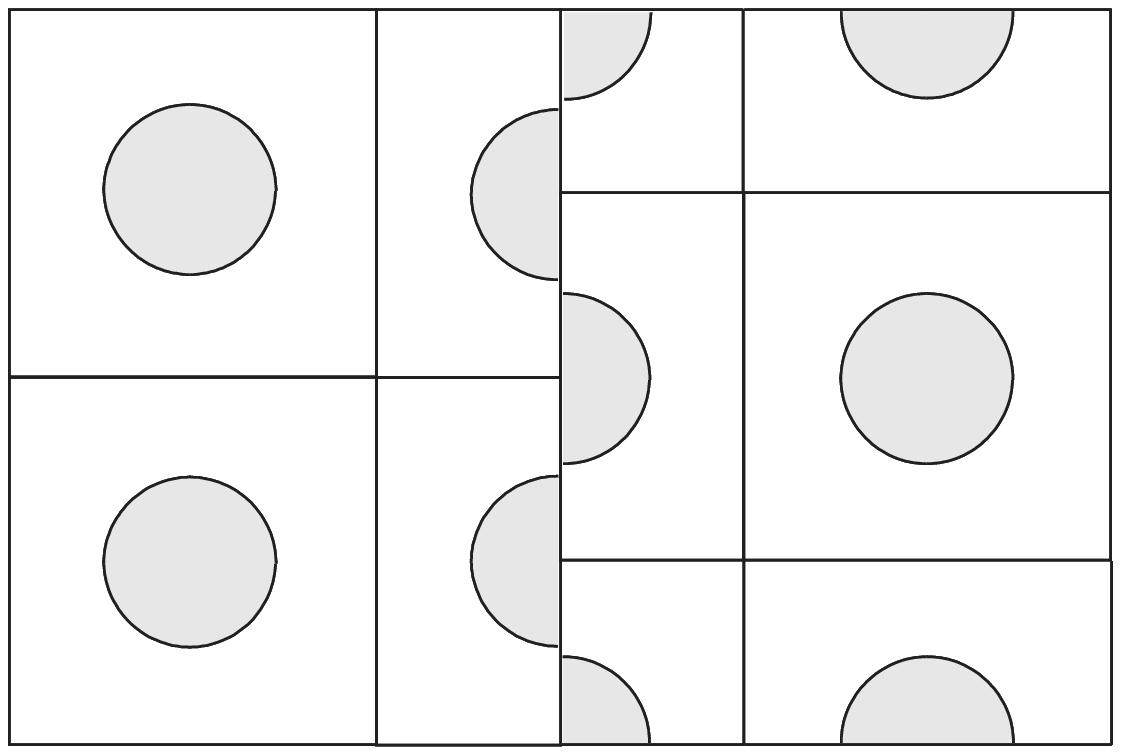}
\caption{Muffin tins: two cases for dislocation in the
$y$-direction.}
\end{center}
\end{figure}

 Note that, if we had chosen $x_0 = 1/2$, then nothing at all
 would have happened for translation in the $y$-direction.

 (The authors thank A.~Ruschhaupt, Hannover, for asking about translation
 in the $y$-direction.)
\section{Appendix: continuity and regularity of eigenvalues}
In this appendix, we discuss several basic facts concerning
continuity and regularity of the eigenvalue branches for the
one-dimensional dislocation problem. We first consider potentials
$V$  from the class $\P \subset L_{1,\text{loc,unif}}(\R)$ as in
(2.1) where the eigenvalues are continuous functions of the
dislocation parameter $t$. In the subsequent estimates we will use
$$
  \norm{V}_{1,\text{loc,unif}} := \sup_{y \in \R} \int_y^{y+1}
  |V(x)| \d x
 \eqno{(A.1)}
$$
as a natural norm on $L_{1,\text{loc,unif}}(\R)$. As is well known
(cf., e.g., [CFrKS]), any potential $V \in
L_{1,\text{loc,unif}}(\R)$ is relatively form-bounded with respect
to $h_0$ with relative form-bound zero. More precisely, we have
the following lemma.\\[.25cm]
{\bf A.1.~Lemma.} {\it For any $\eps > 0$ there exists a constant
$C_\eps \ge 0$ such that for any $V \in L_{1,\text{\rm
loc,unif}}(\R)$ we have
$$
  \int_\R |V| \, |\phi|^2 \d x
   \le \norm{V}_{1,\text{\rm loc,unif}}\,\left(\eps \norm{\phi'}^2 +
   C_\eps \norm{\phi}^2 \right), \qquad \phi \in \mathcal H^1(\R).
  \eqno{(A.2)}
$$
}
\vspace{-.5cm} \proof For $f\in\Ccinf(\R)$ with support contained
in $(0,\eps)$ we have $\infnorm{f}\leq\sqrt\eps\norm{f'}$. Let
$(\zeta_n)_{n\in\N}$ denote a (locally finite) partition of unity
on the real line with the properties:
$\supp\zeta_1\subset(0,\eps)$, each $\zeta_n$ is a translate of
$\zeta_1$, $M:=\sup_{x\in\R}\sum_{n\in\N}|\zeta_{n}'(x)|^2$ is
finite and $\sum_{n\in\N}\zeta_{n}^2(x)=1$ for all $x\in\R$. By
the IMS localization formula (see [CFrKS]), we have for any
$\varphi\in\Ccinf(\R)$,
$$\norm{\varphi'}^2=\ska{-\varphi''}{\varphi}=\sum_{n=1}^{\infty}\norm{(\zeta_n\varphi)'}^2
-\sum_{n=1}^{\infty}\norm{\zeta_n'\varphi}^2\geq\sum_{n=1}^{\infty}\norm{(\zeta_n\varphi)'}^2-M\norm{\varphi}^2,$$
so that
\begin{align}
  \int|V(x)||\varphi(x)|^2\dx
  & \le\sum_{n=1}^{\infty}\infnorm{\zeta_n\varphi}^2\int_{\supp\zeta_n}|V(x)|\dx \nonumber\\
  &  \leq \eps\,\left(\norm{\varphi'}^2+M\norm{\varphi}^2\right)
       \,\norm{V}_{1,\text{\rm loc,unif}}. \nonumber
\end{align}
The general case follows by approximation and Fatou's lemma.
\endproof
For $V \in \P$, the function
$$
   \vartheta_V(s) := \int_0^1 |V(x + s) - V(x)| \d x, \qquad 0\le s \le 1,
   \eqno{(A.3)}
$$
is continuous and $\theta_V(s) \to 0$, as $s \to 0$. Furthermore,
for $W_t$ is as in eqn.~(1.1), we have $\norm{W_t - W_{t'}}_{1,
\text{loc,unif}} = \theta_V(t - t')$. This leads to the following
lemma.
\vskip1em\noindent {\bf A.2.~Lemma.} {\it Let $V \in \P$, $E_0 \in
\R \setminus \sigma(H_{t_0})$, and write $\eps_0 := \text{\rm
dist}(E_0, \sigma(H_{t_0}))$. Then there is $\tau_0 > 0$ such that
$H_t$ has no spectrum in $(E_0-\eps_0/2, E_0 + \eps_0/2)$ for $|t
- t_0| < \tau_0$. Furthermore, there exists a constant $C \ge 0$
 such that for some $\tau_1 \in
(0,\tau_0)$
$$
  \norm{(H_t - E_0)^{-1} - (H_{t_0} - E_0)^{-1}} \le C \theta_V(t - t_0),
   \qquad |t - t_0| < \tau_1.
  \eqno{(A.4)}
 $$
}
\vspace{-.5cm}\proof Without loss of generality we may assume that
$V \ge 1$. Let ${\bold h}_t$ denote the quadratic form associated
with $H_t$. Applying Lemma A.1 (with $\eps := 1$) we see that
$$
   \left|{\bold h}_{t_0}[u] - {\bold h}_t[u] \right|
   \le \int_\R |W_t - W_{t_0}| \, |u|^2 \d x
   \le C_1 \theta_V(t - t_0) {\bold h}_{t_0} [u], \qquad
   u \in {\mathcal H}^1(\R),
$$
with some constant $C_1$. The desired result now follows by [K;
Thm.~VI-3.9].
\endproof
\vskip.25em We therefore see that $H_{t_n} \to H_{t_0}$ in the
sense of norm resolvent convergence if  $t_0 \in [0,1]$,
$(t_n)_{n\in\N} \subset [0,1]$ and $t_n \to t_0$. By standard
arguments, this implies that the discrete eigenvalues of $H_t$
depend continuously on $t$. We are now prepared for the proof of
Lemma 2.1.
\vskip1em
{\it Proof of Lemma 2.1.} We consider $t \in \T$, the flat
one-dimensional torus, and we denote the spectral gap by $(a,b)$.
Let $[a',b']\subset(a,b)$.

{(1)} Let $(\eta,\tau) \in (a,b) \times \T$. Since $\sigma(H_\tau)
\cap (a,b)$ is a discrete set, and since $\sigma(H_t)$ depends
continuously on $t$, there is a neighborhood $U_{\eta,\tau}
\subset (a,b) \times \T$ of $(\eta,\tau)$ of the form
$U_{\eta,\tau} = (\eta_1, \eta_2) \times (\tau_1, \tau_2) $
belonging to either of the two following types:
\vskip.5ex {\bf Type (1):} For $\tau_1 < t < \tau_2$ we have
$\sigma(H_t) \cap (\eta_1, \eta_2) = \emptyset$, \vskip.5ex
or
\vskip.5ex {\bf Type (2):} $\eta$ is an eigenvalue of $H_\tau$ and
there is a continuous function $f \colon (\tau_1, \tau_2) \to
(\eta_1, \eta_2)$ such that $f(t)$ is an eigenvalue of $H_t$;
$H_t$ has no further eigenvalues in $(\eta_1,\eta_2)$, for $\tau_1
< t < \tau_2$.
\vskip.5ex
Now the family $\{ U_{\eta,\tau} \mid (\eta,\tau) \in (a,b) \times
\T\}$ is an open cover of $(a,b) \times \T$ and there exists a
finite selection $\{ U_{\eta_i,\tau_i} \}_{i=1,\ldots,N}$ such
that
$$
   [a',b'] \times \T \subset \cup_{i=1}^N  U_{\eta_i,\tau_i}.
$$
As a first consequence, we see that there is at most a finite
number of functions that describe the spectrum of $H_t$ in the
open set $ \cup_{i=1}^N  U_{\eta_i,\tau_i} \supset [a',b'] \times
\T$.

{(2)} Suppose that $(\eta,\tau) \in (a,b) \times \T$ is such that
$\eta \in \sigma(H_\tau)$ and let $f \colon (\tau_1,\tau_2) \to
(\eta_1,\eta_2)$ as above. Consider a sequence $(t_j)_{j\in\N}
\subset (\tau_1,\tau_2)$ with $t_j \to \tau_1$. We can find a
subsequence $(t_{j_k})_{k\in\N}$ such that $f(t_{j_k}) \to {\bar
\eta}$ for some ${\bar\eta} \in [\eta_1,\eta_2]$. It is easy to
see that ${\bar\eta} \in \sigma(H_{\tau_1})$.
 If ${\bar\eta} \in (a,b)$
 the point $({\bar\eta},\tau_1)$ has a neighborhood
 $U_{{\bar\eta},\tau_1}$ of type $(2)$ and we can extend the
domain of definition of $f$ beyond $\tau_1$. It follows that there
exist a maximal open interval $(\alpha,\beta) \subset (0,1)$ and
a (unique)  continuous extension ${\tilde f} \colon (\alpha,\beta)
\to (a,b)$ of $f$ such that ${\tilde f}(t)$ is an eigenvalue of
$H_t$ for all $t \in (\alpha,\beta)$.

{(3)} It remains to show that $\tilde f(t)$ converges to a band
edge as $t \downarrow \alpha$ and as $t \uparrow \beta$.  By the
same argument as above, we find that any sequence $(t_j)_{j\in\N}
\subset (\alpha,\beta)$ satisfying $t_j \to \alpha$ has a
subsequence $(t_{j_k})_{k\in\N}$ such that $\tilde f(t_{j_k}) \to
{\bar\eta}$ for some ${\bar\eta} \in [a,b]$. Here ${\bar\eta}
\notin (a,b)$ because otherwise we could again extend the domain
of definition of ${\tilde f}$ beyond $\alpha$, contradicting the
maximality property of the interval
 $(\alpha,\beta)$.

Suppose there are sequences $(t_j)_{j\in\N}, (s_j)_{j\in\N}
\subset (\alpha,\beta)$ such that $t_j \to \alpha$ and $s_j \to
\alpha$ and ${\tilde f}(t_j) \to a$ while ${\tilde f}(s_j) \to b$
as $j \to \infty$. Then for any $\eta' \in (a,b)$ there is a
sequence $(r_j)_{j\in\N} \subset (\alpha,\beta)$ such that $r_j
\to \alpha$ and ${\tilde f}(r_j) \to \eta'$, whence $\eta' \in
\sigma(H_\alpha)$. This would imply that $(a,b) \subset
\sigma(H_\alpha)$, which is impossible. \hfill$\square$
\vskip1.0em
We next turn our attention to the question of Lipschitz-continuity
of the functions $f_j$ in Lemma 2.1. With $\theta_V \colon [0,1]
\to [0,\infty)$ as in (A.3), we study potentials from the classes
$$
\PP_\alpha : = \{ V \in \PP \mid \exists C \ge 0 \colon
\theta_V(s) \le C
 s^\alpha, \forall 0< s \le 1\} ,
  \eqno{(A.5)}
$$
where $0 < \alpha \le 1$. The class $\PP_\alpha$ consists of all
periodic functions $V \in \P$ which are (locally)
$\alpha$-H\"older-continuous in the $L_1$-mean; for $\alpha = 1$
this is a Lipschitz-condition in the $L_1$-mean. The class $\PP_1$
is of particular practical importance since it contains the
periodic step functions. It will be shown below that $\P_1$
coincides with the class of periodic functions on the real line
which are locally of bounded variation. We first prove
Lipschitz-continuity of the eigenvalues of $H_t$ for $V \in \P_1$.

\vskip1em\noindent {\bf A.3.~Proposition.} {\it For $V \in \P_1$,
let $(a,b)$ denote any of the gaps $\Gamma_k$ of $H$ and let $f_j
\colon (\alpha_j, \beta_j) \to (a,b)$ be as in Lemma 2.1. Then the
functions $f_j$ are uniformly Lipschitz-continuous. More
precisely, for each gap $\Gamma_k$ there exists a constant $C_k\ge
0$ such that for all $j$
$$
  |f_j(t) - f_j(t')| \le C_k |t - t'|, \qquad \alpha_j \le t, t' \le \beta_j.
$$
}
\vspace{-.5cm} \proof As in the proof of Lemma 2.6 we can find a
finite number of levels $E_1, \ldots, E_\ell \in (a,b)$ and a
partition $0=\tau_0 < \tau_1 < \ldots < \tau_{\ell-1} < \tau_\ell
= 1$ such that $E_j \notin \sigma(H_t)$ for all $t \in I_j :=
[\tau_{j-1}, \tau_j]$ and for $j=1, \ldots, \ell$. Now $V \in
\P_1$ implies $\norm{W_t - W_{t'}}_{1,\text{loc,unif}} =
\theta_V(t-t') \le C |t-t'|$ and we conclude with the aid of Lemma
A.2 that there are constants $c_1, \ldots, c_\ell \ge 0$ such that
$$
   \norm{(H_t - E_j)^{-1} - (H_{t'} - E_j)^{-1}} \le c_j |t-t'|,
   \qquad t, t' \in I_j.
$$
This implies that the min-max-values $\mu_k(s)$ of $(H_s -
E_j)^{-1}$ satisfy
$$
   |\mu_k(t) - \mu_k(t')| \le c_j |t - t'|,
    \qquad t, t' \in I_j.
$$
By the spectral mapping theorem, the eigenvalues of $H_t$ in
$(E_j, b)$ are in bijection with the eigenvalues of $(H_t -
E_j)^{-1}$ in $(\frac{1}{b-E_j},\infty)$. We now let $C :=
\max\{c_1, \ldots, c_\ell\}$ to finish our proof.
\endproof
\vskip.5em\noindent
\noindent{\bf A.4.~Remarks.}

(a) By the same argument, we obtain the following result on
H\"older-continuity: If $0<\alpha <1$ and $V \in \P_\alpha$, then
each of the functions $f_j \colon (\alpha_j,\beta_j) \to (a,b)$ is
locally uniformly H\"older-continuous (as defined in [GT]), i.e.,
for any compact subset $[\alpha_j', \beta_j'] \subset (\alpha_j,
\beta_j)$ there is a constant $C = C(j, \alpha_j', \beta_j')$ such
that $|f_j(t) - f_j(t')| \le C |t - t'|^\alpha$, for all $t, t'
\in [\alpha_j', \beta_j']$. Note that our method does not
necessarily yield a uniform constant for the whole interval
$(\alpha_j, \beta_j)$, much less a constant that would be uniform
for all $j$.
\vskip.5ex
(b)  For analytic potentials $V$, it is shown in [K1] that the
eigenvalue branches $f_j$ in Lemma 2.1 depend analytically on $t$.
This is a simple consequence of the fact that, for real analytic
$V$, the $H_t$ form a holomorphic family of self-adjoint operators
in the sense of Kato. In [K2], the author  proves that the $f_j$
are squares of $W^1_2$-functions near the gap edges if the
potential is in $L_2(\T)$.

\vskip1em

We finally give a characterization of the class $\P_1$.

\vskip1em\noindent {\bf A.5.~Proposition.} {\it Let $BV_{\text{\rm
loc}}(\R)$ denote the space of real-valued functions which are of
bounded variation over any compact subset of the real line.

Then $\P_1 = \P \cap BV_{\text{\rm loc}}(\R)$.
     }
\vskip1em
It is easy to see that any $V \in \P \cap BV_{\text{\rm loc}}(\R)$
belongs to $\P_1$: certainly, any $V \in \P$ which is monotonic
over $[0,1]$ is an element of $\P_1$ and any function of bounded
variation can be written as the difference of two monotonic
functions.

The converse direction is established by the following result due
to J.\ Voigt, Dresden; cf.\ also [EG, Chapter 5] for related
material on $BV$-functions of several variables.
\vskip1em\noindent {\bf A.6.~Lemma.} {\it Let $f \in
L_{1,\text{\rm loc}}\,(\R,\R)$ be periodic with period $1$ and
suppose that there are $c \ge 0$, $\eps > 0$ such that
$$
    \int_0^1 |f(x+t) - f(x)| \d x \le c t, \qquad \forall 0 < t < \eps.
   \eqno{(A.6)}
$$
Consider $f$ as a function in $L_1(\T)$, with $\T$ denoting the
one-dimensional torus.

We then have: the distributional derivative $\partial f$ is a
(signed) Borel-measure $\mu$ on $\T$ and there is a number $a \in
\R$ such that $f(x) = a + \mu([0,x))$, a.e. in $[0,1) \simeq \T$.
In particular, $f$ has a left-continuous representative of bounded
variation. } \vskip1em

\noindent{\it Proof.} Defining $\eta \colon C^1(\T) \to \R$ by
$$\eta(\phi) := - \int_0^1 \phi' f \d x,$$ we may compute
\begin{align}
   - \int_0^1 \phi' f \d x
 & = \lim_{t \to 0} \int_0^1 \frac{1}{t} (\phi(x-t) - \phi(x)) f(x) \d x
 \nonumber\\
 &  =  \lim_{t \to 0} \int_0^1 \phi(x) \frac{1}{t} (f(x+t) -
             f(x)) \d x,
 \nonumber
\end{align}
and the assumption (A.6) yields the estimate $|\eta(\phi)| \le c
\norm{\phi}_\infty$. Since $C^1(\T)$ is dense in $C(\T)$, the
functional $\eta$ has a unique continuous extension to all of
$C(\T)$; we denote the extension by the same symbol $\eta$. By the
Riesz representation theorem there is a measure $\mu$ such that
$\eta(\phi) = \int \phi \d \mu$ for all $\phi \in C(\T)$.
Furthermore, for $\phi \in C^1(\T)$ we have $ - \int_0^1 \phi' f
\d x= \int_0^1 \phi \d \mu$, and we see that $\mu = \partial f$ on
$\T$ in the distributional sense.
The choice $\phi := 1$ yields $\int_\T \d \mu = - \int_0^1 \phi' f
\d x = 0$ and the function ${\tilde f}(x) := \mu([0,x))$ satisfies
$\partial {\tilde f} = \mu$. This is easy to check: for $\phi \in
C^1(\T)$ we have
\begin{align}
 \int {\tilde f} \phi' \d x
  & = \int_0^1 \int_{0 \le y < x} \d \mu(y) \phi'(x) \d x
  \nonumber\\
  &  = \int_{0 \le y < 1} \int_y^1 \phi'(x) \d x \d \mu(y)
    = - \int_{[0,1)} \phi(y) \d \mu(y).
  \nonumber
\end{align}
We therefore see that $\partial (f - {\tilde f}) = 0$; hence there
is some $a$ such that $f - {\tilde f} = a$. \hfill$\square$

\end{document}